# Security for IEEE P1451.1.6-based Sensor Networks for IoT Applications

Hiroaki Nishi
Faculty of Science and Technology, Keio University
Yokohama, Japan
west@ keio.jp

Janaka Wijekoon
Keio University
Yokohama, Japan
janaka@west.sd.keio.ac.jp

Eugene Y. Song
Communications Technology Laboratory
National Institute of Standards and Technology
Gaithersburg, MD, USA
eugene.song@nist.gov

Kang B. Lee
IEEE Life Fellow
Gaithersburg, MD, USA
Kang.Lee@ieee.org

*Abstract*— There are many challenges for Internet of Things (IoT) sensor networks including the lack of robust standards, diverse wireline and wireless connectivity, interoperability, security, and privacy. Addressing these challenges, the Institute of Electrical and Electronics Engineers (IEEE) P1451.0 standard defines network services, transducer services, transducer electronic data sheets (TEDS) format, and a security framework to achieve sensor data security and interoperability for IoT applications. This paper proposes a security solution for IEEE P1451.1.6-based sensor networks for IoT applications utilizing the security framework defined in IEEE P1451.0. The proposed solution includes an architecture, a security policy with six security levels, security standards, and security TEDS. Further, this paper introduces a new service to update access control lists (ACLs) to regulate the access for topic names by the applications and provides an implementation of the security TEDS for IEEE P1451.1.6-based sensor networks. The paper also illustrates how to access security TEDS that contain metadata on security standards to achieve sensor data security and interoperability.

I. INTRODUCTION

Sensors and sensor networks are poised for large-scale adoption in the heterogeneous Internet of Things (IoT) and industrial IoT (IIoT) applications. However, due to the lack of standardization in IoT sensors and sensor networks [1], diverse connectivity, interoperability, security, and privacy have become major challenges. Since sensors are small size and low-cost, have low-power consumption, and are constrained in computation and communication resources, security is a unique and critical challenge for them. Cyberattacks are targeting IoT sensors and sensor networks in more significant numbers, in more industries, and with greater sophistication than ever before, creating barriers for their ubiquitous use in IoT/IIoT applications [1]. Therefore, a security framework and standardization are essential to provide countermeasures against cyberattacks in sensor networks [2].

The Institute of Electrical and Electronics Engineers (IEEE) 1451 family of standards for smart transducer interfaces for sensors and actuators defines specifications for transducer and network services and interfaces. As shown in Fig. 1, all these standard interfaces are based on P1451.0 that is a core standard of the IEEE 1451 family of standards. P1451.0 defines common functions and global identities of applications (APP) (1451 client), network-capable application processor (NCAP) (1451 server), transducer interface module (TIM) (1451 sensors and/or actuators), network services messaging between APP and NCAP, transducer services messaging between NCAP and TIM, security framework, time synchronization framework, and transducer electronic data sheets (TEDS) [3]. Those standards designated with a "P" in Fig.1 are ongoing standards development projects for either developing new standards or revising existing standards.

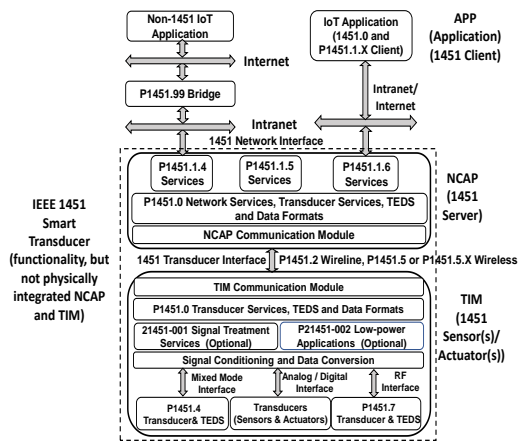

Fig. 1. The architecture of IEEE 1451 family of standards

IEEE P1451.1.6 defines a method for transporting P1451.0 network service messages over a message queuing telemetry transport (MQTT) network to establish a lightweight, messaging protocol for sensor network communications [4]. This paper proposes a security solution for IEEE P1451.1.6 standard-based sensor networks and is organized as follows. Related works are described in Section II. Section III describes the proposed security solution for IEEE P1451.1.6 sensor networks using MQTT. The security TEDS implementation of IEEE P1451.1.6 sensor networks is provided in Section IV. Section V provides the summary and conclusion.

II. RELATED WORKS

Among a few studies about security for IEEE 1451 standards-based smart sensors and sensor networks, Wu et al.

proposed a cross-layer security mechanism that deals with the requirements of authentication, integrity, confidentiality, and availability across the communication process in IEEE 1451 smart transducers [5]. Similarly, D. James et al. discussed some security aspects of sensor networks by integrating simple network management protocol (SNMP) management information base (MIB) with IEEE 1451.4 TEDS [6]. IEEE 1451 optional security TEDS is proposed and discussed in [7], and it includes a signature, public key, encryption, and hashing algorithms used for landmark deployment and security. The proposed security TEDS described in [8] can also be used for secure key exchange mechanisms and permission management. Rocha studied and discussed sensor data protection of secure communications of low-power and wireless sensors that are energy harvested with microbial fuel cells [9]. A security framework for IEEE P1451.0-based IoT sensor networks is proposed and described, including three security policies and six security levels to meet security requirements [3] [10]. It defines security TEDS to describe security protocol information for IEEE P1451.0-based sensor networks.

There are a few studies discussing implementing IEEE P1451.0 and P1451.1.6 sensor networks. Among them, Carratù et al. have demonstrated the feasibility of a wireless sensor network based on low-cost hardware and using the MQTT communication protocol according to the IEEE 1451.1.6 standard [11]. Nishi and Lee provided a reference implementation of IEEE P1451.0 and P1451.1.6 sensor networks using MQTT to verify IEEE P1451.0 network service messaging [12]. Nishi et al. applied IEEE P1451.0 and P1451.1.6 standards-based sensor networks for temperature monitoring and airflow control to balance and improve greenhouse environments [13]. Unfortunately, none of these studies discussed the security of these reference implementations. Therefore, this paper focuses on a security solution for IEEE P1451.1.6 sensor networks based on the IEEE P1451.0 security framework.

## III. SECURITY SOLUTION FOR IEEE P1451.1.6-BASED SENSOR NETWORKS

### A. Architecture of Secure IEEE P1451.1.6 Sensor Networks

Fig. 2 shows an architecture of secure IEEE P1451.1.6 standards-based sensor networks for IoT applications [3] [5], which includes two-level securities: wide-area network (WAN) security and wireless local area network (WLAN) security. As shown in Fig. 2, a secure P1451.1.6 standards-based WAN consists of a few APPs (1451 clients) and a set of NCAPs (1451 server) or IEEE 1451 smart transducers (e.g., a combination of NCAP and TIM). These NCAPs are connected to the APPs via the Internet/intranet (e.g., Ethernet or cellular). The secure communications between APPs and NCAPs are based on IEEE P1451.0 network services and IEEE P1451.1.6 MQTT interfaces and security standards. Also, a secure IEEE P1451.0 and P1451.5.X standards-based WLAN consists of an NCAP (1451 server) and a set of wireless TIMs (WTIMs). These WTIMs are connected with the NCAP via wireless mediums defined in IEEE 1451.5 and P1451.5.X interfaces. The secure communications between the NCAP and WTIMs are based on IEEE P1451.0 transducer services and IEEE 1451.5, P1451.5.X interfaces and security standards.

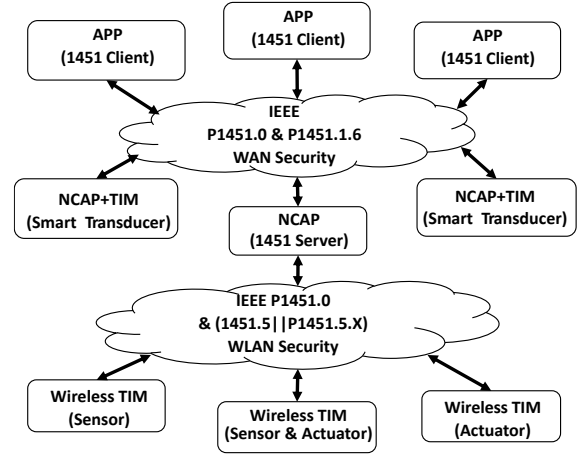

Fig. 2. Architecture for secure IEEE P1451.1.6 standards-based sensor networks.

### B. Security for IEEE P1451.1.6 Standards Sensor Networks

IEEE P1451.0 defines the security framework for IEEE 1451 standards-based sensor networks, including security requirements, policy, security levels, security standards, and security TEDS [3]. Adhering to IEEE P1451.0, IEEE P1451.1.6 security requirements, policy, levels, and TEDS formats shall be specified and listed as follows.

**Security Requirements:** MQTT defines the security requirements: "MQTT Client and Server implementations SHOULD offer Authentication, Authorization, and secure communication options" [14]. P1451.1.6 should provide secure mechanisms as follows:
- Authentication of clients and servers;
- Authorization of access to server resources;
- Integrity of MQTT control packets and application data contained therein; and,
- Privacy of MQTT control packets and application data contained therein.

Applications concerned with critical infrastructure, personally identifiable information, or other personal or sensitive information are strongly advised to use these security capabilities [14].

**Security Policy:** Security policies defined in IEEE P1451.0 are divided into three main categories: encryption, authentication, and authorization. Encryption is the process of encrypting encoded data or information. In accordance with the OASIS MQTT standard, P1451.1.6 advanced encryption standard (AES) and ChaCha20 are recommended for encryption in sensor networks. Authentication includes identity verification and integrity-checking mechanisms. The connect packet of MQTT contains a username and password for the simple authentication of an MQTT broker. The OASIS MQTT standard introduces examples of its use for lightweight directory access protocol (LDAP) or open authorization (OAuth). Authorization determines and enforces the access rights of P1451.1.6 NCAP and APP, allowing the entity to use specific resources. In P1451.1.6, a client should be authorized to use the client identifier to access the MQTT broker session states. Moreover, the client should request access control permissions to the MQTT-ACS to publish topics or to subscribe using topic filters [14].

**Security Levels:** Security levels of IEEE P1451.1.6 sensor networks are specified in P1451.0 [3]. Each security level is a combination of the three security policies except for level N, which means no security.
- **Level A:** supports only encryption to keep IEEE P1451.1.6 sensor network data encrypted.
- **Level B:** only supports authentication to control access based on the identity of P1451.1.6 NCAP and APP.
- **Level C:** combines encryption and authentication.
- **Level D:** combines authentication and authorization.
- **Level E:** combines all policies: encryption, authentication, and authorization.

Security levels of IEEE P1451.1.6 sensor networks mean different combinations of three security policies, which does not mean that one level is more or less secure than others.

**Security of P1451.0 and P1451.1.6 WAN interface:**

The network stack of IEEE 1451.0 and P1451.1.6 WAN includes a physical layer (e.g., Cellular, long-term evolution (LTE), 5G, Ethernet, virtual private network (VPN), Intranet), transport layer (e.g., transmission control protocol (TCP) and user datagram protocol (UDP)), and application layer (e.g., IEEE P1451.1.6 and P1451.0). Security requirements and recommendations for the IEEE P1451.0 and P1451.1.6 WAN interface include that:
- The integrity of P1451.1.6 NCAP and APPs configuration settings can be verified;
- The integrity of P1451.1.6 NCAP and APPs software can be verified;
- The identities of P1451.1.6 NCAPs and APPs can be enforced; and,
- Authentication of P1451.1.6 NCAP and APP software can be enforced.

*C. Security Standards of IEEE P1451.1.6 MQTT*

MQTT is a client-server publish/subscribe messaging transport protocol. It is lightweight, open, simple, and designed to be easy to implement [14]. MQTT targets tiny devices with limited computational, memory, and power resources. IEEE P1451.1.6 defines a method for transporting IEEE 1451.0 network service messages over a network using MQTT to establish a lightweight, simplified protocol structure to handle the IEEE P1451.1.X WAN communications between APP and NCAP [4].

Table I. List of security standards of IEEE P1451.1.6.

| No. | Security Standard Name | Description |
|---|---|---|
| 1 | NIST Cyber Security Framework (CSF) | The NIST Cybersecurity Framework (CSF):<br>CSF V1.1:<br>https://nvlpubs.nist.gov/nistpubs/CSWP/NIST.CSWP.04162018.pdf<br>CSF V2.0:<br>https://nvlpubs.nist.gov/nistpubs/CSWP/NIST.CSWP.29.ipd.pdf |
| 2 | PCI-DSS | Payment Card Industry Data Security Standard (PCI-DSS):<br>https://www.pcisecuritystandards.org/pci_security/ |
| 3 | FIPS-140-2 | Security Requirements for Cryptographic Modules (FIPS PUB 140-2):<br>https://csrc.nist.gov/csrc/media/publications/fips/140/2/final/documents/fips1402.pdf |
| 4 | NSA Suite B | NSA Suite B Cryptography:<br>http://www.nsa.gov/ia/programs/suiteb_cryptography/ |
| 5 | ChaCha20 | ChaCha20 and Poly1305 for IETF Protocols:<br>https://tools.ietf.org/html/rfc7539 |
| 6 | AES | Advanced Encryption Standard (AES) (FIPS PUB 197):<br>https://csrc.nist.gov/csrc/media/publications/fips/197/final/documents/fips-197.pdf |
| 7 | ISO 29192 | ISO/IEC 29192-1:2012 Information technology -- Security techniques -- Lightweight cryptography -- Part 1: General:<br>https://www.iso.org/standard/56425.html |
| 8 | LDAP | Sermersheim, J., Ed., "Lightweight Directory Access Protocol (LDAP): The Protocol", RFC 4511, DOI 10.17487/RFC4511, June 2006. http://www.rfc-editor.org/info/rfc4511 |
| 9 | OAuth | [RFC6749] Hardt, D., Ed., "The OAuth 2.0 Authorization Framework", RFC 6749, DOI 10.17487/RFC6749, October 2012, http://www.rfc-editor.org/info/rfc6749. |
| 10 | TLS | [RFC5246] Dierks, T., and E. Rescorla, "The Transport Layer Security (TLS) Protocol Version 1.2", RFC 5246, DOI 10.17487/RFC5246, August 2008, http://www.rfc-editor.org/info/rfc5246. |
| 11 | VPN | Virtual Private Network (VPN) for authentication |
| 12 | Username/Password | MQTT Authentication: Clients send username and password to the MQTT broker. |
| 13 | Client Identifier | MQTT Authentication: Client Identifier- each MQTT client has a unique client identifier. |
| 14-127 | Reserved | |
| 128 | User-defined | MQTT-ACL |
| 129-255 | User-defined | |

MQTT security standards are summarized and listed in Table I [14]. Table I shows the enumeration No., the security standard name, and a description of each security standard in P1451.6. Note that the enumeration number is used to encode the standard as an integer in packet headers and does not indicate a priority or ranking of the different standards. The industry security standards include the National Institute of Standards and Technology (NIST) CSF, Payment Card Industry Data Security Standard (PCI-DSS), Federal Information Processing Standard-140-2 (FIPS-140-2), and National Security Agency (NSA) Suite B. Both ChaCha20 and AES are used for lightweight cryptography in constrained devices. The International Organization for Standardization (ISO) 29192 makes recommendations for cryptographic primitives specifically tuned to perform on constrained "low-end" devices. The authentication of clients by the server includes LDAP or OAuth, a VPN for authentication, and Transport Layer Security (TLS) certificate. Both username/password and client identifier are also used for authentication. Authorization may be based on information provided by the client, such as username, the hostname/IP address of the client, or the outcome of authentication mechanisms. The authorization of clients by the server includes a TLS certificate and VPN. The integrity of application messages and MQTT control packets includes both a TLS hash algorithm and VPN to verify the integrity of data. The privacy of application messages and MQTT control packets include TLS for the encryption of data sent over a network, VPN for the privacy of data access, and AES for the encryption of application messages.

*D. Security TEDS*

TEDS are used to describe the characteristics of the transducers in IEEE 1451 sensor networks. The security TEDS are defined in the IEEE P1451.0 standards [3] and are accessed using a read or write TEDS command. Security TEDS indicate how many security protocols are employed through the *NumOfProtocols* entry and provide detailed information on each protocol by the *SecurityStdName* entry and *SecurityStdVersion* entry. The security TEDS of the IEEE P1451.1.6-based sensor networks should be based on the definitions given in IEEE P1451.0 standards [3], and security levels and security standards should be specified in IEEE P1451.1.6.

*E. MQTT Access Control Lists*

When considering MQTT, in addition to secure communication using login and password supported by TLS, the use of hardening the access control settings is also vital. Access control regulates the names of topics that each user can access and their access privileges, thus enabling fine-grained access control. For example, well-known MQTT broker services, such as Mosquitto**, provide an access control lists (ACL) method. However, in general, only the administrator of the broker's server can change the ACL settings. This is because the change always requires access rights to the host where the MQTT broker runs, as well as access rights to the access control file of Mosquitto, and security control level permissions, such as sending a SIGHUP using root privileges to re-read the access control list. Moreover, all these access rights must be held and executed by the MQTT broker host. Therefore, it makes them less convenient and more costly to operate.

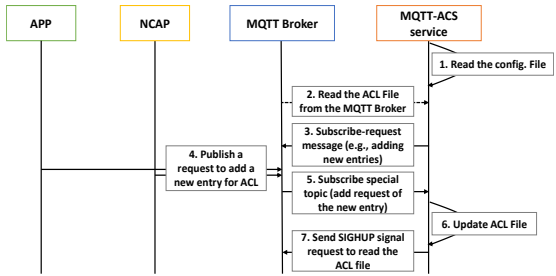

Fig. 3 Access control update for IEEE P1451.1.6.

MQTT access permissions can be made more convenient by allowing only services as MQTT clients with special permissions to set and change the ACL entries. As illustrated in Fig. 3, a new service, MQTT-ACL Client Service (MQTT-ACS), has been established to secure such authority transfers; MQTT-ACS runs concurrently with the MQTT broker on the server on which the MQTT broker runs. The details of the MQTT-ACS service illustrated in Fig. 3 are as follows. Firstly, the service reads the configuration file and starts subscribing to a specific topic (e.g., adding new entries). The configuration file can provide the topic name, such as '1451.1.6/ACL/CONFIG'. The access tokens are shared between the service and APPs or NCAPs, which have the right to control the ACS. For example, when the NCAP wants to change the ACL according to the update of security TEDS, the NCAP publishes an ACL update message to the topic, and the MQTT Broker forwards that message to the MQTT-ACS service. Upon receiving the message, the MQTT-ACS service processes the request and updates the ACL file; accordingly, the service issues a SIGHUP signal to the MQTT broker to read the file. Finally, the ACL is updated, and a new access policy becomes active.

Only MQTT-ACS clients, authorized by using a pre-defined login and password and holding pre-defined access tokens, can modify access control list using commands defined by MQTT-ACS. The MQTT-ACS client may use the commands specified by MQTT-ACS to control access. It also describes which MQTT-ACS clients can define access control separately and to what extent. Generally, it is defined so that access rights can only be controlled at a level lower than the topic handled by the client in queries. Note that this restriction can also be freely controlled.

## IV. IMPLEMENTATION OF SECURITY TEDS OF IEEE P1451.1.6-BASED SENSOR NETWORKS

*A. Security Use Case of IEEE P1451.1.6 Sensor Networks*

Fig. 4 shows a security use case for the IEEE P1451.1.6 sensor network consisting of one APP (client 1) and one NCAP (client 2). This use case mainly focuses on security of the WAN instead of WLAN shown in Fig. 2. Based on the P1451.1.6 security configuration, the APP can securely publish a command to the NCAP to read security TEDS via the MQTT Mosquitto broker. The NCAP can securely subscribe the command from the APP via the MQTT broker, and then can securely publish the reply to the APP via the broker based on the command received. In the end, the APP can securely subscribe the corresponding reply from the NCAP based the command sent.

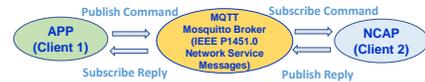

Fig. 4. Security use case for P1451.1.6 sensor networks.

*B. Implementation of Security TEDS of IEEE P1451.1.6 Sensor Networks*

Table II shows an example of IEEE P1451.1.6 security TEDS in the use case of Fig. 4. In this example, security TEDS Id includes 1 and 6 (P1451.1.6 standard), 16 (security

TEDS access code), 2 (P1451.0 version 2), 1 (tuple length). The security level is E, a combination of encryption, authentication, and authorization. Security standards include TLS V1.3, login/password V1.0, and MQTT-ACL V1.0 with their SecurityStdNames cross-referenced against Table I. Table III shows the different version options available for the TLS standards.

Table II. Example of IEEE P1451.1.6 security TEDS.

| No. | Field name | Description | Data type | # Octets | Value | Description |
|---|---|---|---|---|---|---|
| — | | Length | UInt32 | 4 | | |
| 0-2 | — | Reserved | — | — | | |
| 3 | TEDSID | TEDS Identification Header | UInt8Array | 5 | 1 6 16 2 1 | (1 6): 1451.1.6 16: security TEDS 2: TEDS version (2) 1: Tuple length |
| 4-9 | — | Reserved | — | — | | |
| | | Security related information | | | | |
| 10 | Level | Security level | Octet | 1 | E | Level E |
| 11 | NumOfStandards | Number of security standards | UInt8 | 1 | 3 | 3 standards |
| 12 | SecurityStdName1 | Security standard name 1 | UInt8 | 1 | 10 | TLS |
| 13 | SecurityStdVersion1 | Security standard version 1 | UInt8 | 1 | 4 | V1.3 |
| 14 | SecurityStdName2 | Security standard name 2 | UInt8 | 1 | 12 | Login/password |
| 15 | SecurityStdVersion2 | Security standard version 2 | UInt8 | 1 | 1 | V1.0 |
| 16 | SecurityStdName3 | Security standard name 3 | UInt8 | 1 | 128 | MQTT-ACL |
| 17 | SecurityStdVersion3 | Security standard version 3 | UInt8 | 1 | 1 | V1.0 |
| — | | Checksum | UInt16 | 2 | | |

Table III. Versions of TLS standards.

| No. | Version | Description |
|---|---|---|
| 0 | Default | |
| 1 | TLS 1.0 | TLS 1.0, RFC 8996, 1999 |
| 2 | TLS 1.1 | TLS 1.1, RFC 8996, 2006 |
| 3 | TLS 1.2 | TLS 1.2, RFC 5246, 2008 |
| 4 | TLS 1.3 | TLS 1.3, RFC 8446, 2018 |
| 5-255 | Open for manufacturers | |

Table IV. Read security TEDS command & reply messages.

| Read security TEDS message | | Type | Name | Value |
|---|---|---|---|---|
| Command Message | Header | NetSvcType<br>NetSvcId<br>MsgType<br>UInt16 | netSvcType<br>netSvcId<br>msgType<br>msgLength | 3<br>2<br>1<br>? |
| | Body | UUID<br>UUID<br>UUID<br>UInt16<br>TedsAccessCode<br>UInt32<br>TimeDuration | appId<br>ncapId<br>timId<br>channelId<br>tedsAccessCode<br>tedsOffset<br>timeout | ?<br>?<br>?<br>?<br>16<br>0<br>2s |
| Reply Message | Header | NetSvcType<br>NetSvcId<br>MsgType<br>UInt16 | netSvcType<br>netSvcId<br>msgType<br>msgLength | 3<br>2<br>2<br>? |
| | Body | UInt16<br>UUID<br>UUID<br>UUID<br>UInt16<br>UInt32<br>OctetArray | errorCode<br>appId<br>ncapId<br>timid<br>channelId<br>tedsOffset<br>rawTEDSBlock | 0<br>?<br>?<br>0<br>0<br>0<br>? |

Table IV shows both the read TEDS command and reply messages defined in P1451.0 that are exchanged in Fig. 4 [3]. Both command and reply messages consist of a message header and a message body. The message header includes network service type (netSvcType), network service Id (netSvcId), message type (msgType) and message length (msgLength). Refer to the values specified in Table II for the read TEDS command. The contents of the security TEDS in Table II should be included in the rawTEDSBlock of the reply message body. The network service message of read security TEDS is created by using P1451.0 (D0) command and D0 message modes of P1451.1.6, and the responses were given in D0 TEDS and D0C TEDS [13], respectively.

Fig. 5 shows the sequence diagram of read security TEDS. The APP can read the security TEDS by sending the P1451.0 Read TEDS command to the NCAP via the MQTT broker; then, the NCAP can reply by sending the security TEDS back to the APP.

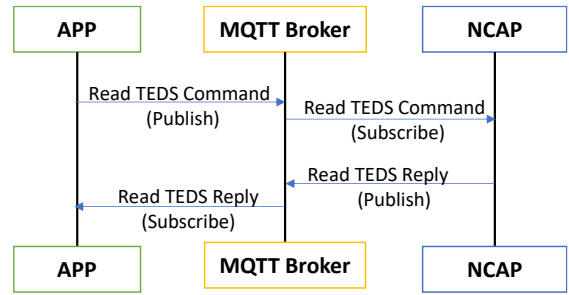

Fig. 5. Sequence diagram of read security TEDS.

### C. Experiment Setup and Result of Security TEDS of IEEE P1451.1.6 Sensor Networks

These NCAP functions with security TEDS are implemented on the reference model of [12] as shown in Fig. 6. Both NCAP and APP have been updated to support security TEDS. Experiment setup is described as follows:

- Two NCAPs are implemented on NodeMCU and Raspberry Pi V4, respectively.
- APP and V4 MQTT-ACL service are implemented on another PC of M1T+.
- Mosquitto, an MQTT broker, is executed on the micro PC of GMKtec NucBox 5. MQTT broker and MQTT-ACL are running on Ubuntu 24.04.

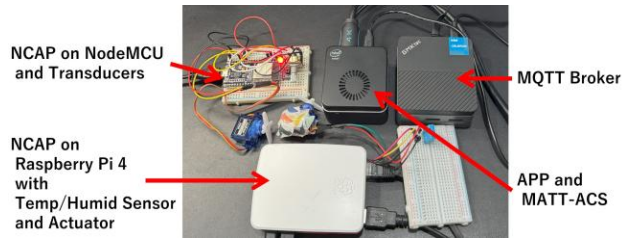

Fig. 6. Implementation of test environment.

Fig. 7 shows the P1451.1.6 security TEDS screenshot of the implementation. The security TEDS includes security levels E, number of security standards (3), and the names and versions of three security standards including TCL (10), username/password (12) and MQTT-ACL (128).

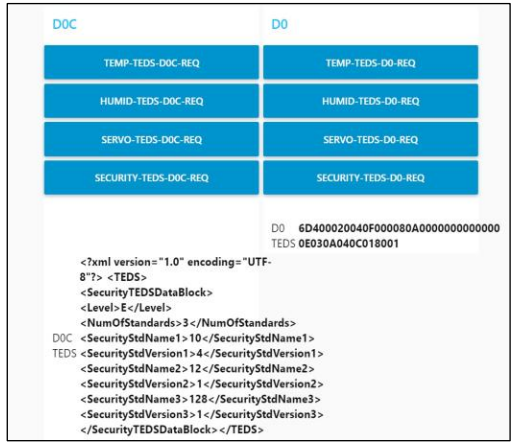

Fig. 7. Result of read security TEDS.

## V. CONCLUSION

This paper describes a security solution for IEEE P1451.1.6-based sensor networks, including architecture, security policy and six security levels, security standards, and security TEDS. A list of security standards for IEEE P1451.1.6 MQTT is described in this paper, and a new service, MQTT-ACS, has been introduced to update ACLs to regulate the access for names and topics by each APP. The implementation of the security TEDS of IEEE P1451.1.6 sensor networks is also provided in the paper to validate P1451.0 security TEDS access. The operation of the APP, which reads the security TEDS information from the NCAP by sending a command request for the security TEDS to the NCAP via the IEEE P1451.1.6 MQTT network interface, was explained. Moreover, the security TEDS information of the IEEE P1451.1.6 security protocol between the APP and NCAP is also described in the paper.

ACKNOWLEDGMENT

The sensors portion of this work were supported by MOE, Japan, "Technology Development and Demonstration Project for Regional Symbiosis and Cross-Sectoral Carbon Neutrality (Second Round)/Development and demonstration of a ZEG for contributing to the decarbonization of horticulture". Moreover, the authors express their gratitude for the support of hardware design to JST CREST Grant Number JPMJCR19K1 and for the standardization processes to the CSTI, SIP, the 3rd period of SIP, Smart energy management system, Grant Number JPJ012207 (Funding agency: JST).

The authors gratefully acknowledge the valuable comments and suggestions of Thomas Roth for supporting the work.